\documentclass[12pt,a4paper,notitlepage]{article}
\usepackage{caption}
\usepackage{authblk}
\usepackage{graphicx}
\usepackage{caption}
\usepackage{subcaption}
\usepackage{cite}
\usepackage{enumerate}
\usepackage{amsmath}
\usepackage[top=2cm, bottom=2cm, left=2cm, right=2cm]{geometry}

\date{\small \today}

\title{Containment and resolution of hadronic showers at the FCC}

\author[a]{\small Tancredi Carli}
\author[a]{\small Clement Helsens}
\author[a]{\small Ana Henriques Correia}
\author[a]{\small Carlos Solans S\'anchez}
\affil[a]{\small CERN}

\begin{document}

\maketitle

\abstract{
The particles produced at a potential Future Circular Collider with $\sqrt{s}$ = 100 TeV are of unprecented energies.
In this document we present the hadronic shower containment and resolution parametrizations based on Geant4 simulations 
for the Hadronic calorimetry needed for conceptual detector design at this energy.
The Geant4 toolkit along with FTFP\_BERT physics list are used in this study. 
Comparisons are made with test-beam data from the ATLAS Tile hadronic calorimeter.
These simulations motivate a 12~$\lambda$ calorimeter in order to contain at 98\% level TeV single hadron showers 
and multi-TeV jets and keep a pion energy resolution constant term of approximately 3\%.
}

\section{Motivation}
\label{sec:Motivation}

With a center of mass energy of $\sqrt{s}$ = 100~TeV a completely new energy regime is opening and 
thus opportunities for discovering physics beyond the Standard Model (SM).
New heavy particles such as extra-gauge bosons might decay hadronically, leading to final states 
in the detector with two very high $p_T$ jets.
At those energies, the resolution of the hadronic calorimeter is dominated by its constant term (as shown 
in Equation~\ref{eq:reso}) consequently the longitudinal leakage of hadronic showers will play a major role in the calorimeter resolution.
In order to measure precisely these objects, the calorimeter requirements with respect to the LHC experiments, 
have to be reconsidered. 

\begin{figure} [!ht]
\centering 
\includegraphics[width=0.48\textwidth]{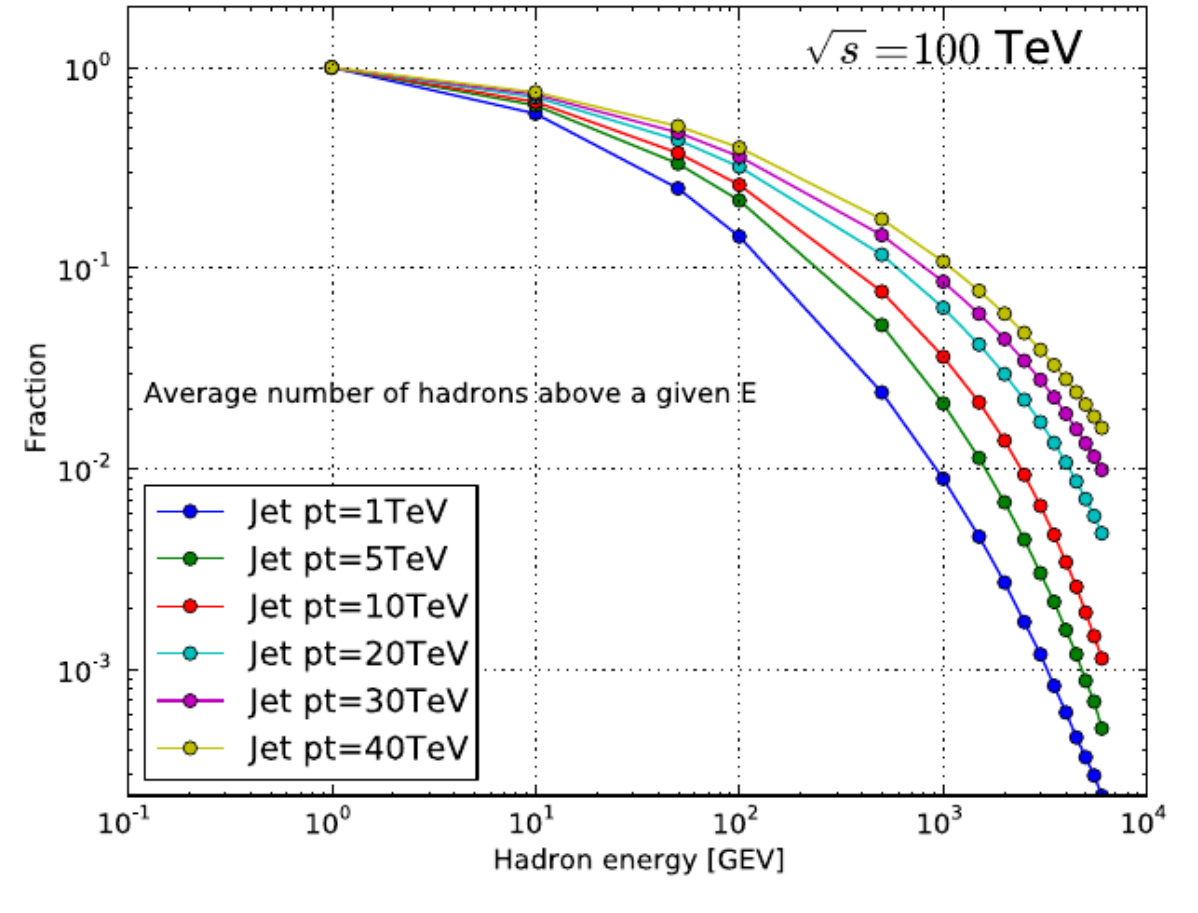}
\includegraphics[width=0.48\textwidth]{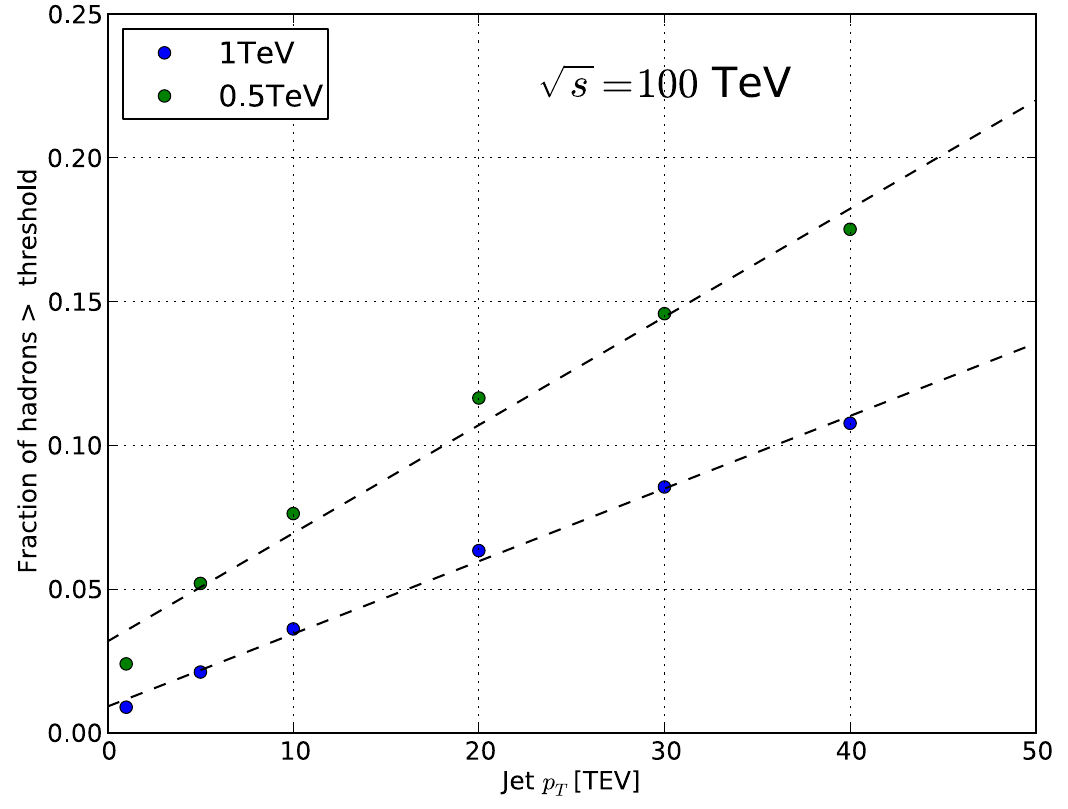}
\caption{Fraction of hadrons above an energy threshold versus the hadron energy (left) and versus the jet $p_T$ (right) 
from a MC simulation at $\sqrt{s}$ = 100 TeV.\label{fig:FracHadrons}}
\end{figure}

As shown in Figure~\ref{fig:FracHadrons}, the fraction of high energy hadrons~\footnote{hadrons are defined as the particles used to build the jets excluding photons} 
(above 1~TeV) in jets, increases significantly with jet $p_T$. 
Table~\ref{table:HadronsInJets} shows the average number of single hadrons above an energy threshold for different jet $p_T$ values. 
For example, a 30~TeV jet is in average composed of 62 hadrons, 9.1 (15\%) of them having an energy above 1~TeV.

The note is naturally organized as follows. The details of the simulation are introduced in Section~\ref{sec:Simulation}, 
followed by the study of the longitudinal containment of single hadrons and jets in Sections~\ref{sec:Containment} and \ref{sec:JetContainment} respectively. 
Finally, pion resolution for different calorimeter depths are derived in Section~\ref{sec:Resolution} and conclusions are drawn in Section~\ref{sec:Conclusions}.

\begin{table}[!ht]
\centering
\begin{tabular}{|c|cccccc|}
\hline
Jet $p_{T}$ [TeV] & All & $>$50 GeV & $>$500 GeV & $>$1 TeV & $>$2 TeV & $>$5 TeV \\
\hline
 1 &  34 &  8.5 &  4.9 &  0.8 & 0.16 & 0.01 \\
 5 &  52 & 17.3 & 11.3 &  2.7 & 0.6  & 0.05 \\
10 &  59 & 22.1 & 15.3 &  4.5 & 1.3  & 0.11 \\  
20 &  62 & 27   & 19.8 &  7.2 & 2.6  & 0.44 \\ 
30 &  62 & 29.5 & 22.4 &  9.1 & 3.7  & 0.83 \\
40 &  60 & 30.5 & 23.8 & 10.5 & 4.6  & 1.24 \\
\hline
\end{tabular}
\caption{Average number of hadrons above a threshold for various jet $p_{T}$.\label{table:HadronsInJets}}
\end{table}

\section{Simulation details}
\label{sec:Simulation}

\begin{figure}[!h]
\centering
\includegraphics[width=10cm]{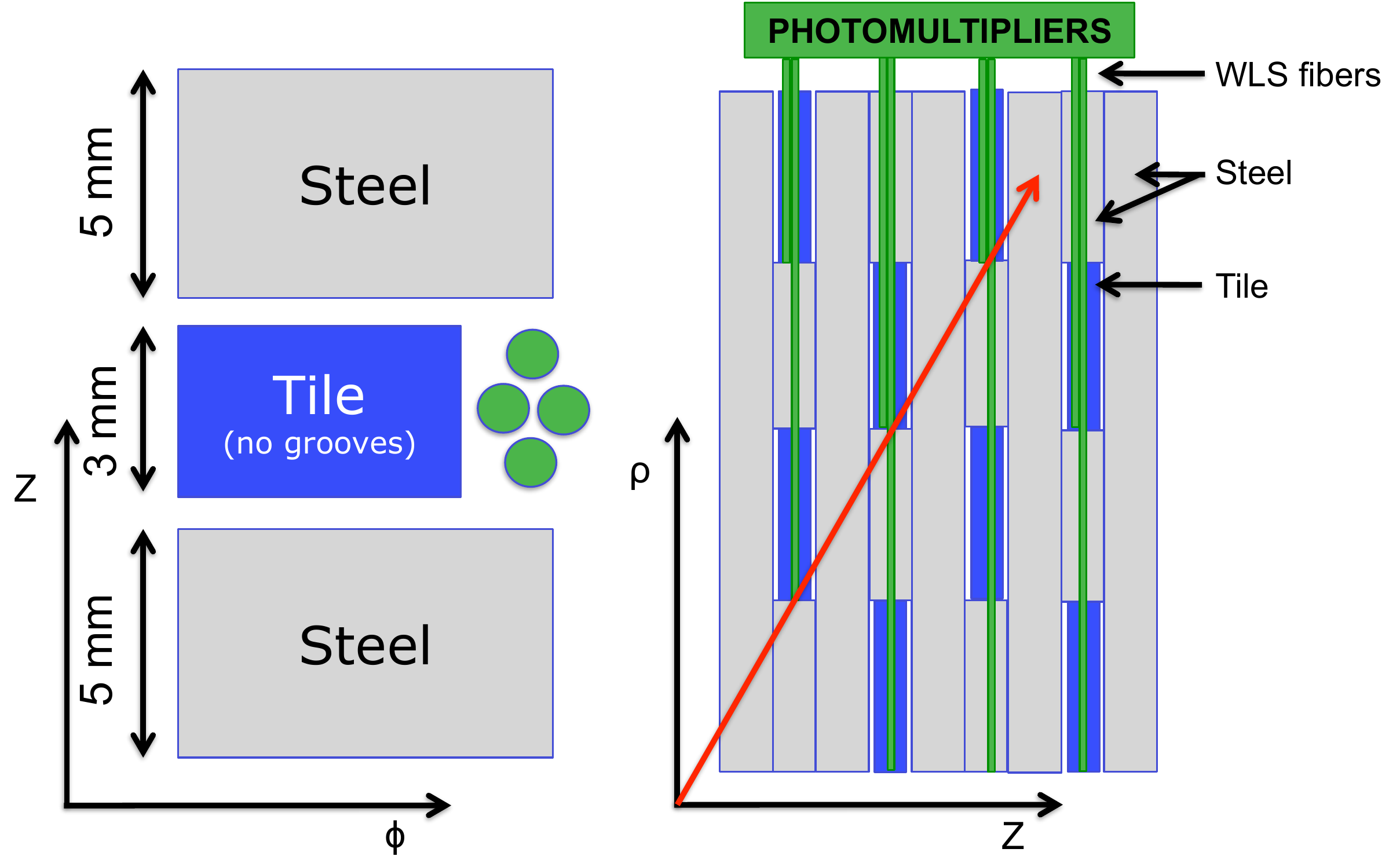}
\caption{Illustration of the Tile hadronic calorimeter geometry in the Z$\phi$ plane and in the $\rho$Z plane used for the simulations. \label{fig:TileGeometry}}
\end{figure}

The simulation used for this study is based on the ATLAS hadronic Tile calorimeter concept \cite{Tilecal} 
in which scintillating tiles are embedded into steel absorber plates placed perpendicular to the traditional orientation in calorimetry, see Figure~\ref{fig:TileGeometry}.
Each tile is read-out by wavelength shifting fibers on both sides and merged into bundles to define cells.
In this study, trapezoidal shaped plates that extend 2$\pi$/128 in the azimuthal direction ($\phi$) are placed along the colliding beam axis direction (Z) 
in a sequence that is repeated 50 times and is made out of four tiles of three different types. 
The first type is a steel plate of 5~mm thickness, the second type is a spacer steel plate 4 mm thick, 
and the third type is a scintillating tile polystyrene based of 3~mm thick (see Figure~\ref{fig:TileGeometry}). 
Each sequence is composed of two master plates separated by a spacer plate followed by an active plate embedded in an air gap of 0.5~mm on each side. 
The total period of the sequence is 18~mm, and the iron to scintillator ratio calculated from the volumes is 4.67 to 1. 

Each tile is 100~mm deep in the perpendicular direction to the beam pipe ($\rho$), and a total of 40 layers are placed in this direction. 
The sequence is altered every layer to alternate the position of the active and spacer plates. 
A 10~mm front steel plate needed at inner radius to ensure the mechanical rigidity of the calorimeter is introduced 
in the MC geometry description since it plays a role in the absorption of low energy incident particles.
With this configuration, the nuclear interaction length of the calorimeter is 20.6~cm, calculated using Bragg's rule \cite{BraggsRule}. 
This calculation is in accordance with previous measurements of the ATLAS Tile calorimeter~\cite{TileInteractionLength}.
The calorimeter used in these simulations is about 20 nuclear interaction lengths deep at a polar angle ($\theta$) measured from Z of $\theta$ = 90$^{\circ}$. 

The simulation software used in this study is Geant4 \cite{Geant4} version 10.1. 
Hadronic inelastic interactions use a tabulation of the Barashenkov pion cross sections and the Axen-Wellisch parameterization of the proton and neutron cross sections. 
The physics list used is known as FTFP\_BERT that uses Bertini-style cascade for hadrons below 5 GeV and FTF (Fritiof) model for interactions of high energy hadrons, above 4 GeV.

Simulation hits are aggregated in each polystyrene (active) tile without any re-weighting on the energy of the active material (raw data).
The energy of each active tile is then stored per event if the total energy deposit is above 1~keV. 
This simulation also takes into account the scintillation non-linearity effects also known as Birk's law \cite{BirksLaw},
for which the total energy deposit at each step of the simulation is weighted by the values extracted from \cite{BirksLawCoefficients}.

\section{Single pion longitudinal containment}
\label{sec:Containment}

The containment for single particles is derived from the simulation of $\pi^{+}$ showers, 
that are produced at the geometrical center of the detector and with an angle of 20$^{\circ}$ 
measured perpendicular to the Z-axis at different kinetic energies ranging from 20 GeV to 10 TeV. 
The energy deposits are measured as a function of nuclear interaction lengths ($\lambda$) 
following the axis of the incident particle ($\overrightarrow{a}$), referred to as depth in the following.
The depth of any given cell in the calorimeter with a position with respect to the geometrical center of the detector, 
denoted by $\overrightarrow{b}$, is measured as the projection to $\overrightarrow{a}$, 
$\lambda = |\overrightarrow{a}| cos(\theta)$, where $\theta$ is the angle between $\overrightarrow{a}$ and $\overrightarrow{b}$. 

\begin{figure}[!ht]
\centering
\includegraphics[width=0.49\textwidth,trim=0.2cm 0 3cm 0,clip]{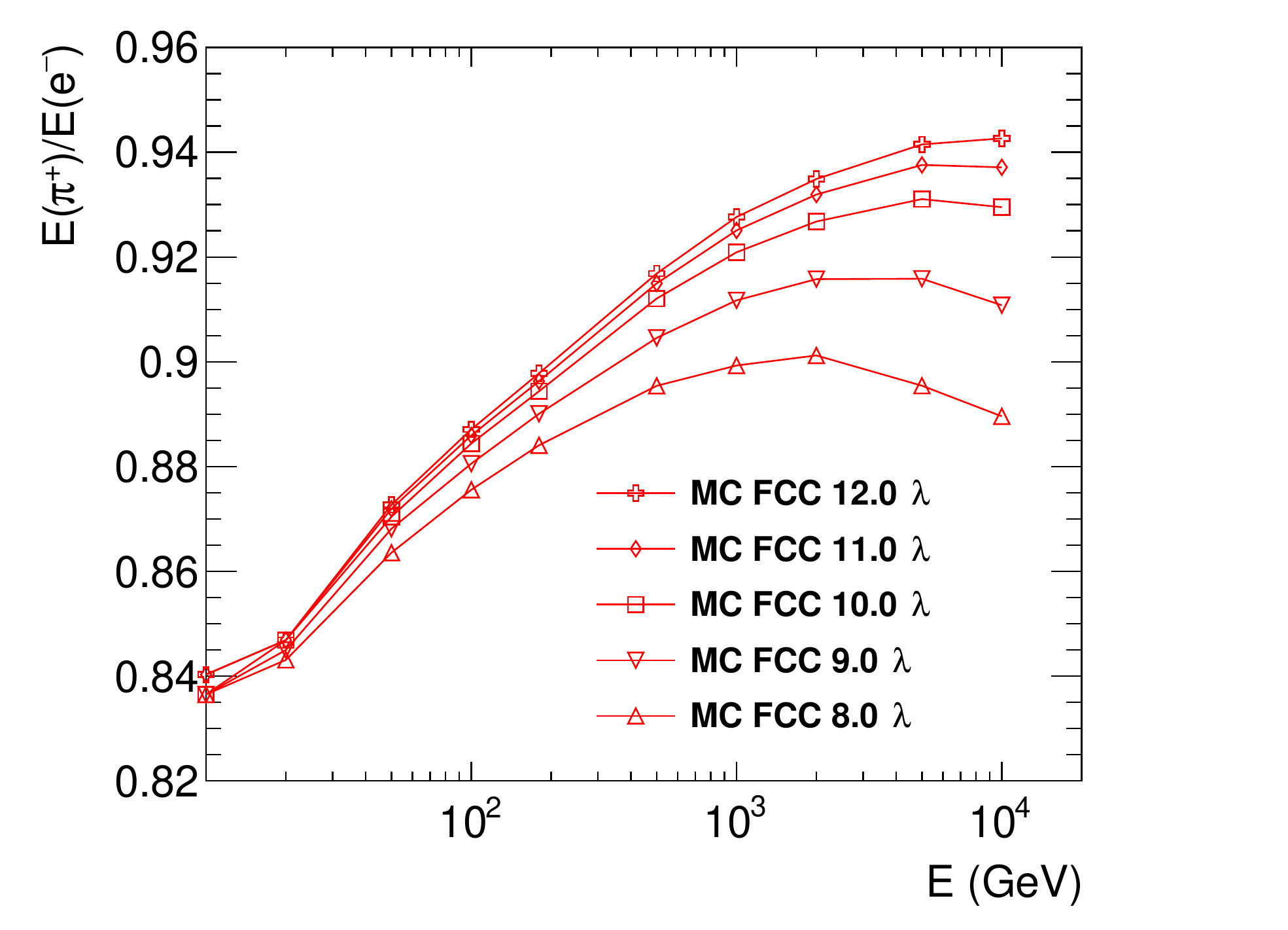}
\includegraphics[width=0.49\textwidth,trim=0.2cm 0 3cm 0,clip]{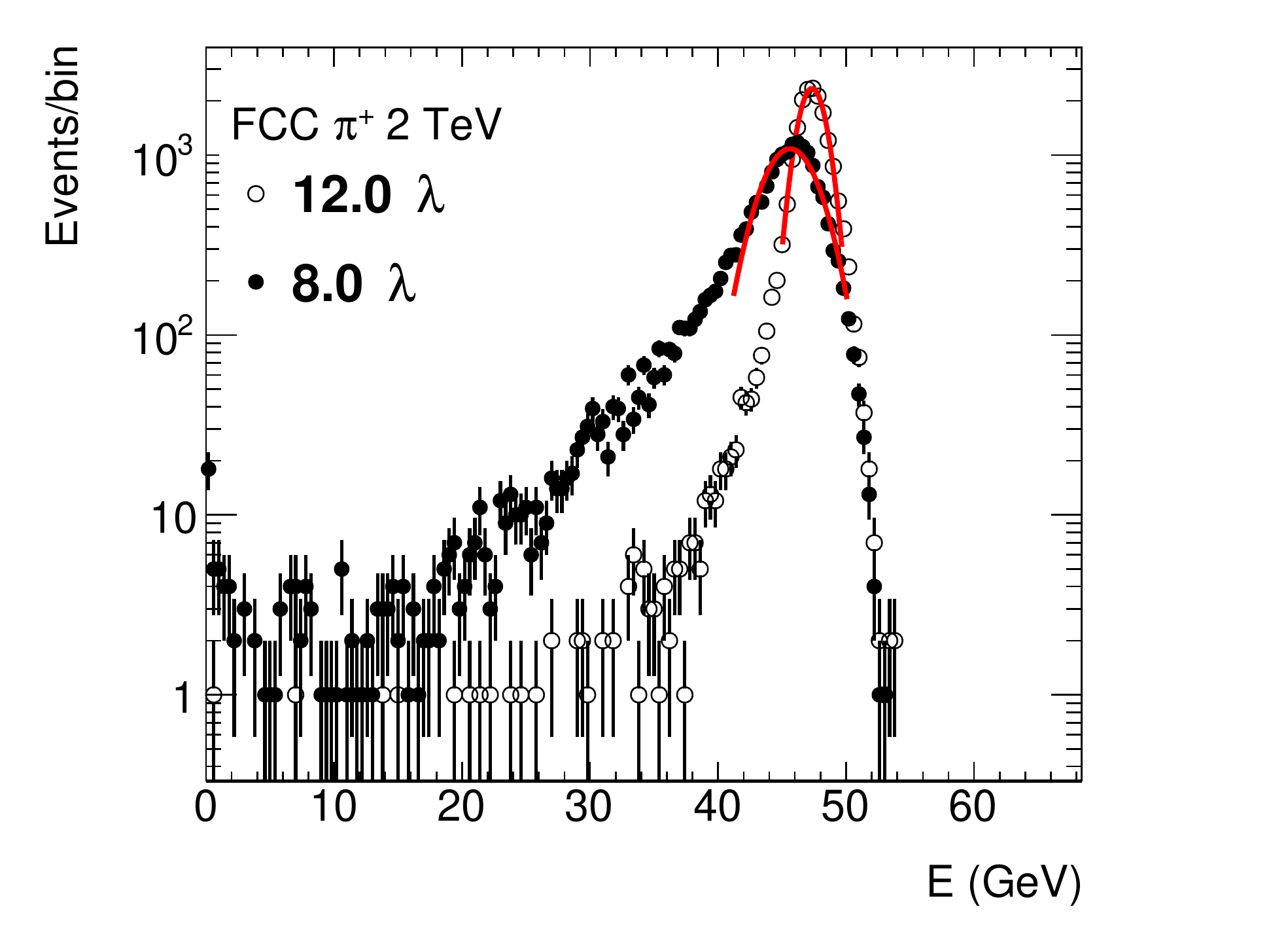}
\caption{Pion response over electron response in the active scintillator volume as a function 
of incident kinetic energy for different calorimeter depths (left) 
and distribution of the deposited energy of a pion with an energy of 2 TeV 
in the active layers of the calorimeter with 8 $\lambda$ and 12 $\lambda$ depths (right).
\label{fig:PiOverEleAndShowerOverlay}}
\end{figure}

The simulated calorimeter set-up is non-compensated with e/h $\sim$ 1.33~\cite{TileTestbeam}.
Consequently the response of the calorimeter to pions, called pion response in the following, is smaller than the response to electrons of the same incident energy and their ratio ($E(\pi^+)/E(e^{-})$) grows with energy. This is shown in Figure~\ref{fig:PiOverEleAndShowerOverlay} (left) and is in accordance with previous measurements~\cite{TileTestbeam}.
Furthermore, smaller calorimeter depths are affected by leakage as the energy of the incident pion increases leading to an extra non-linearity. 
Figure \ref{fig:PiOverEleAndShowerOverlay} (right) shows how low energy tails compare 
between 8 $\lambda$ and 12 $\lambda$ deep calorimeter for a 2 TeV pion. 
The response evaluated as the ratio of the mean of the distributions for 8~$\lambda$ to 12~$\lambda$ is 96~\%.
The size of the low energy tail quantified by the number of events below 3$\sigma$ deviations from the peak, is 11~\% for 8~$\lambda$ and drops down to 3~\% for 12~$\lambda$.

\begin{figure} [!ht]
\centering 
\includegraphics[width=10cm]{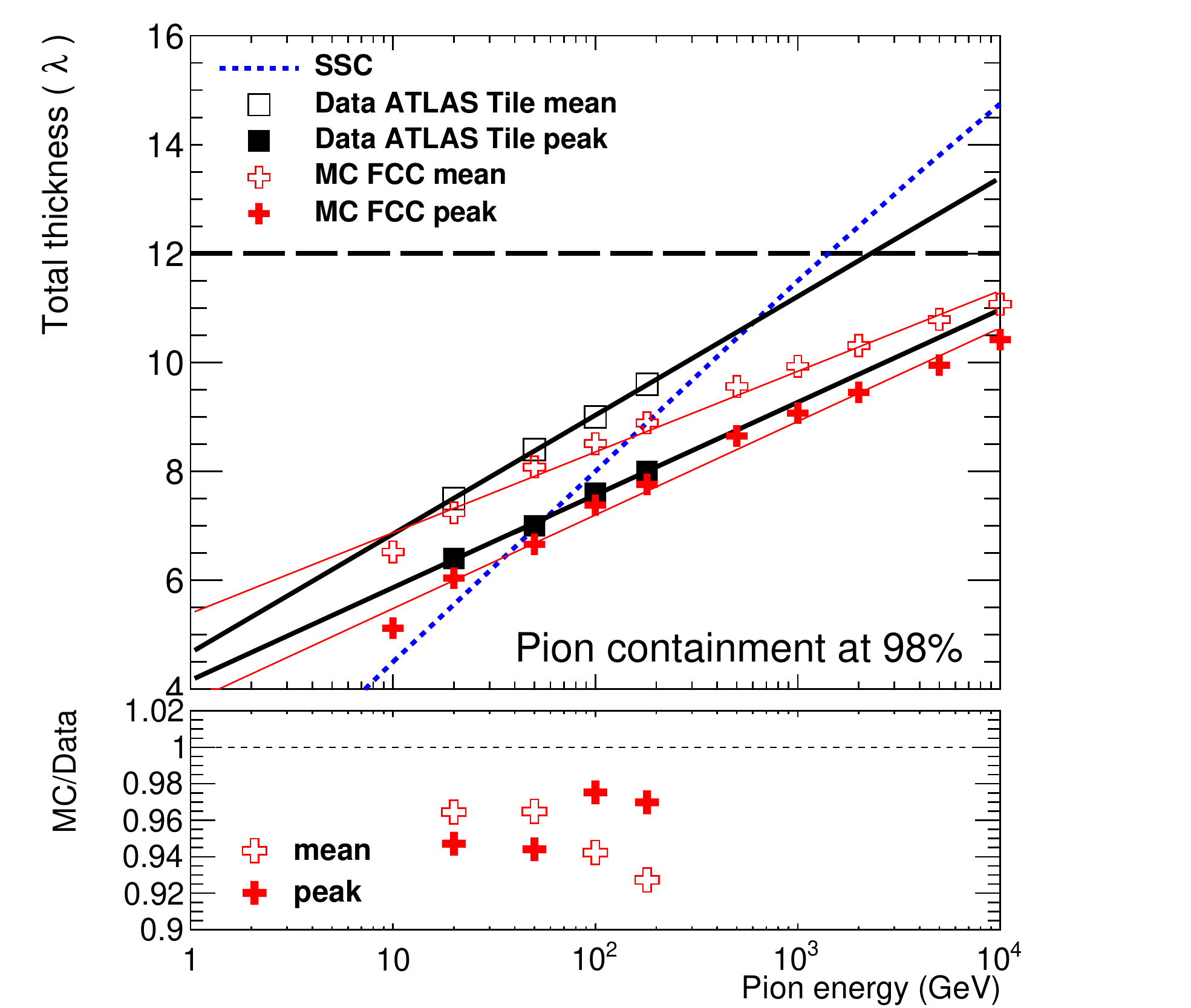}
\caption{Total thickness expressed in nuclear interaction lengths required to contain a single pion ($\pi^+$) up to 98~\% versus the pion incident energy for two different methods explained in the text. SSC parametrization can be found in \cite{SscCalorimetry}. Data corresponds to measurements with the ATLAS Tile calorimeter production modules \cite{Tile20Lambda}.\label{fig:Containment}}
\end{figure}

\begin{table}[!ht]
\centering
\begin{tabular}{|c|cc|cc|}
\hline
Pions  & \multicolumn{2}{c|}{Data} & \multicolumn{2}{c|}{Simulation} \\
\hline
Method & a ($\lambda$/GeV) & b ($\lambda$) & a ($\lambda$/GeV) & b ($\lambda$) \\
\hline
Mean & 0.95 & 4.7 & 0.64 & 5.4 \\
Peak & 0.74 & 4.2 & 0.75 & 3.8 \\
\hline
\end{tabular}
\caption{Values for the parametrization of the total thickness expressed in nuclear interaction lengths required to contain a single pion ($\pi^+$) up to 98~\% as a function of the incident energy of the pion.\label{table:FitParams}}
\end{table}

The containment of the shower is evaluated as the depth  at which a fraction of the incident energy has been absorbed. For this simulation we use containment at 98~\%, which is calculated in two steps. First the total energy absorbed in the calorimeter per event is summed up, and second the depth at which that the cumulative energy is 98~\% of that of the maximum is derived. This method delivers a distribution of depths which is then characterized by the mean value of the distribution and the mean ($\mu$) value of a Gaussian fit between $\pm$2$\sigma$ of the maximum, which are referred to as mean and peak respectively. Figure~\ref{fig:PiOverEleAndShowerOverlay} (right) shows this Gaussian fit.

Figure~\ref{fig:Containment} shows the total thickness needed to absorb 98\% of the energy of a pion versus its incident energy. Simulation is overlaid with the data from ATLAS test-beam measurements 
up to 180~GeV~\cite{Tile20Lambda} for the two methods described. The figure also shows the parametrization for the Superconducting Super-Collider (SSC) derived in 1986 \cite{SscCalorimetry}. 
There is a rather good agreement between data and simulation for the peak method, while the results obtained from the mean of the distribution that accounts for tails due to longitudinal leakage grows slower in MC than in data. This shows that the longitudinal profile of the shower is shorter in MC than in data. The agreement of the simulation to the data is relatively good and within 10 \%.
Data and simulation are parametrized with the function $\lambda = a \cdot ln(E) + b$. The value of the parameters is listed in Table~\ref{table:FitParams}.
Despite the uncertainties, it is clear that more than 10 $\lambda$ are needed to contain few TeV single pions up to 98~\%.

\section{Jet containment}
\label{sec:JetContainment}

Containment of jets is evaluated from single partons hadronized with Pythia 8 \cite{Pythia8}. 
In order to study the reconstruction of isolated partons with approximately fixed $p_T$ and direction, the 
$Z^\prime\rightarrow q\bar{q}$ process in the Les Houches Event (LHE) file format~\cite{Alwall:2006yp} is used.
The LHE files are modified such that the partons $p_T$  from the $Z^\prime$ decay is fixed to a given value in the range [0.2, 10]~TeV, and 
always at the same angle $20^{\circ}$ measured perpendicular to the Z-axis. 
As the $Z^\prime$ is produced at rest, the decay products are back to back.
All particles from Pythia are entered into the Geant4 simulation. 
The  anti-k$_\text{t}$ jet clustering algorithm~\cite{AntiKt} is used with a radius parameter $R = 0.5$. 
The inputs are the energy depositions of the tiles at different depths ($\lambda$) after Geant4 simulation or stable particles entering the Geant4 simulation.
Jets reconstructed from the calorimeter energy deposition or the stable particles are matched by requiring
$\Delta R = \sqrt{\Delta \eta^2 + \Delta\phi^2}  < 0.2$, where $\eta = -ln\tan(\theta/2)$.

\begin{figure} [!ht]
\centering 
\includegraphics[width=10cm]{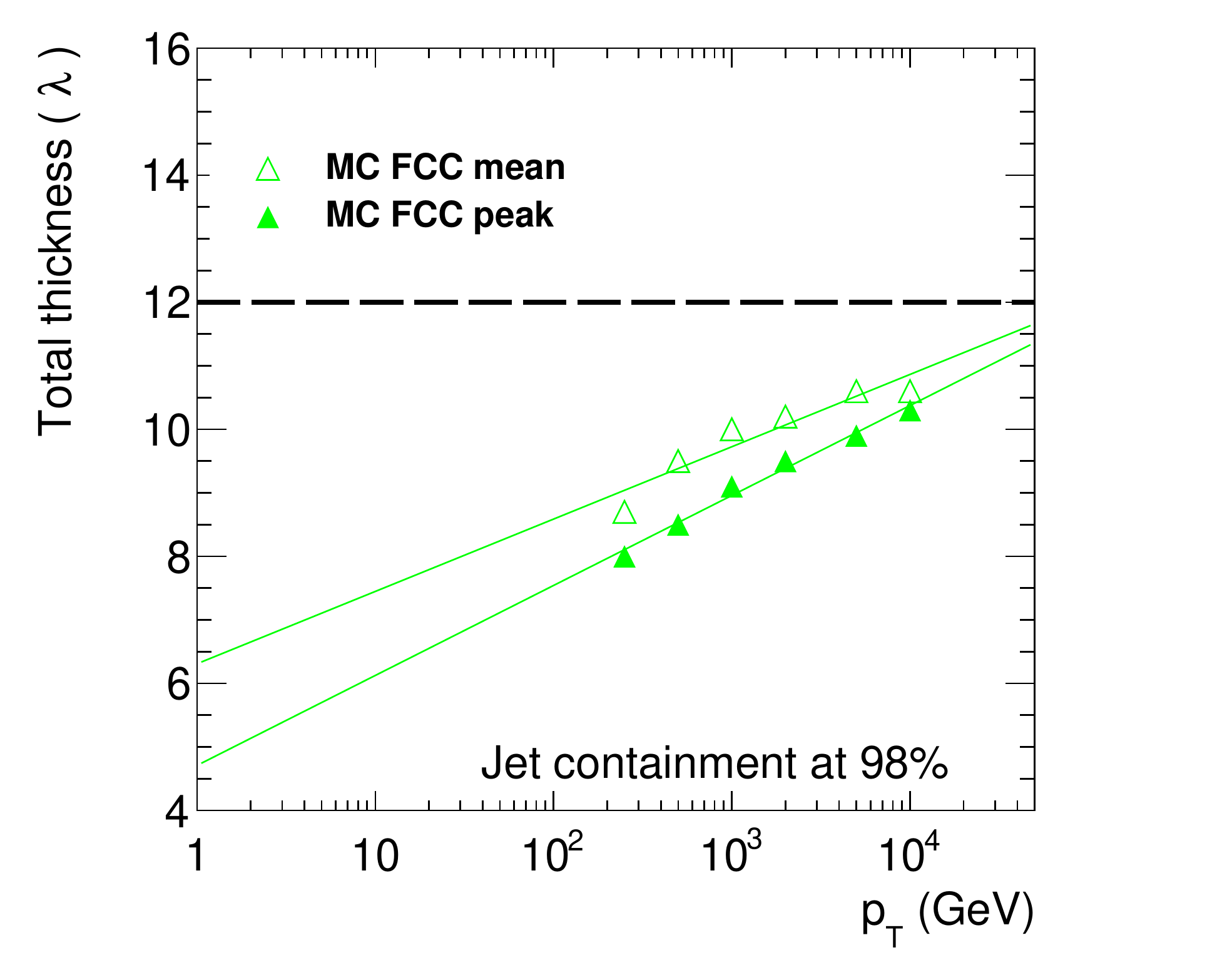}
\caption{Total thickness expressed in nuclear interaction lengths required to contain a jet up to 98~\% vs the $p_T$ of that jet. \label{fig:JetContainment}}
\end{figure}

\begin{table}[!ht]
\centering
\begin{tabular}{|c|cc|}
\hline
Jets   & \multicolumn{2}{c|}{Simulation} \\
\hline
Method & a ($\lambda$/GeV) & b ($\lambda$) \\
\hline
Mean & 0.495 & 6.3 \\
Peak & 0.615 & 4.7 \\
\hline
\end{tabular}
\caption{Values for the parametrization of the total thickness expressed in nuclear interaction lengths required to contain a jet jet up to 98~\% as a function of its $p_T$. \label{table:JetFitParams}}
\end{table}

The particle jet is required to be within 10\% of the $p_T$ of the parton from the $Z^\prime$ decay to avoid selecting events with hard radiations.
Similar energy distributions are obtained at each depth and characterized by the mean and the peak (see Section~\ref{sec:Containment}).
Figure~\ref{fig:JetContainment} shows the depth required to contain 98 \% of the jet energy as a function of $p_T$ of the parton selected as explained above.
The containment scales linearly with the logarithm of the jet $p_T$.
The points can be described by a parameterisation  $\lambda = a \cdot ln(E) + b$ 
where $a$ and $b$ are free parameters. The values are shown in Table~\ref{table:JetFitParams}.
As in the case of single pions, more than 10 $\lambda$ are needed to contain 98\% of the $p_T$ of jets with $p_T$ $>$ 10 TeV. 

\section{Single pion energy resolution}
\label{sec:Resolution}
\begin{figure} [!ht]
 \centering 
 \includegraphics[width=0.49\textwidth,trim={0.3cm 0 3.1cm 0},clip]{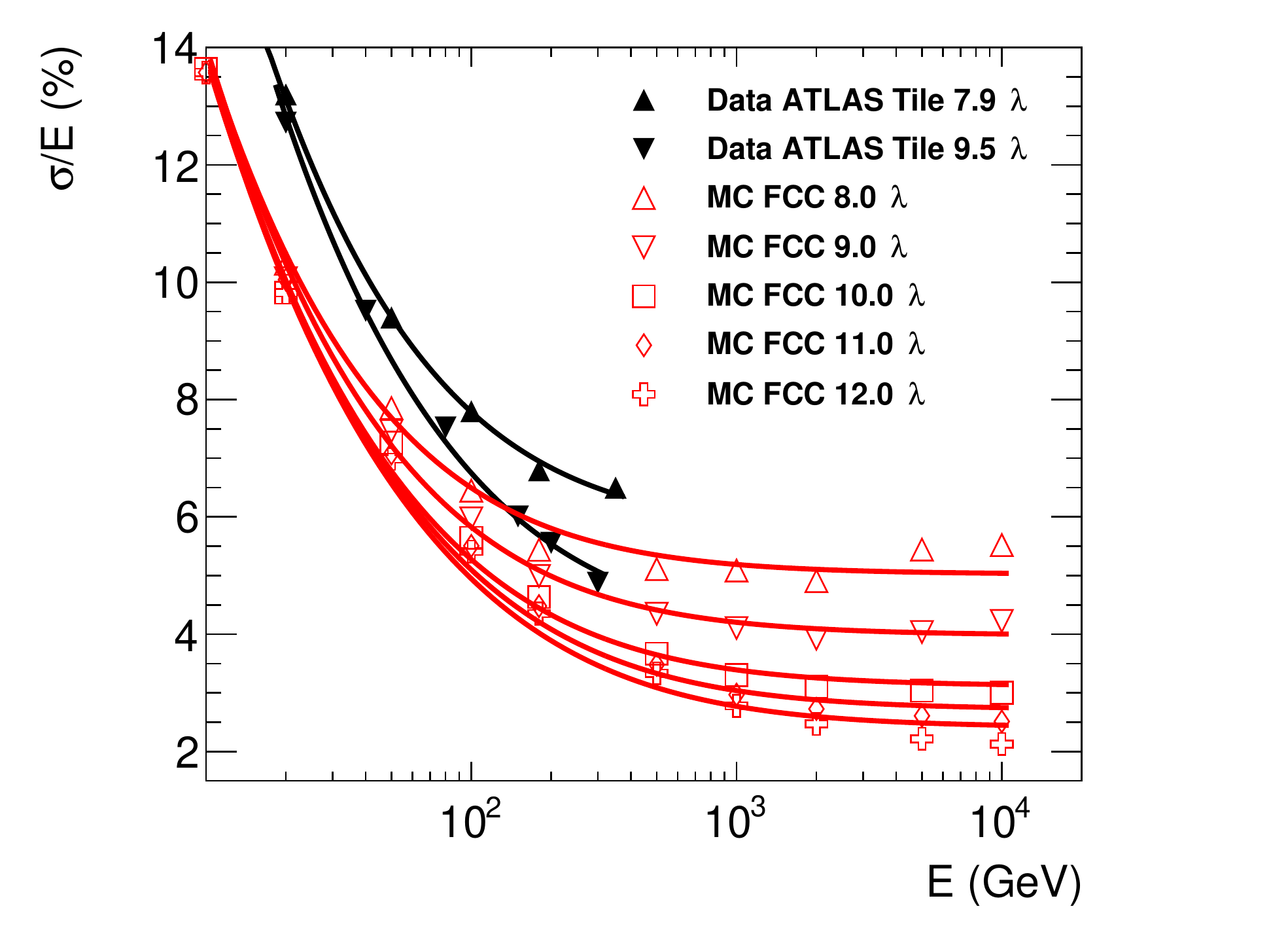}
 \includegraphics[width=0.49\textwidth,trim={0.3cm 0 3.1cm 0},clip]{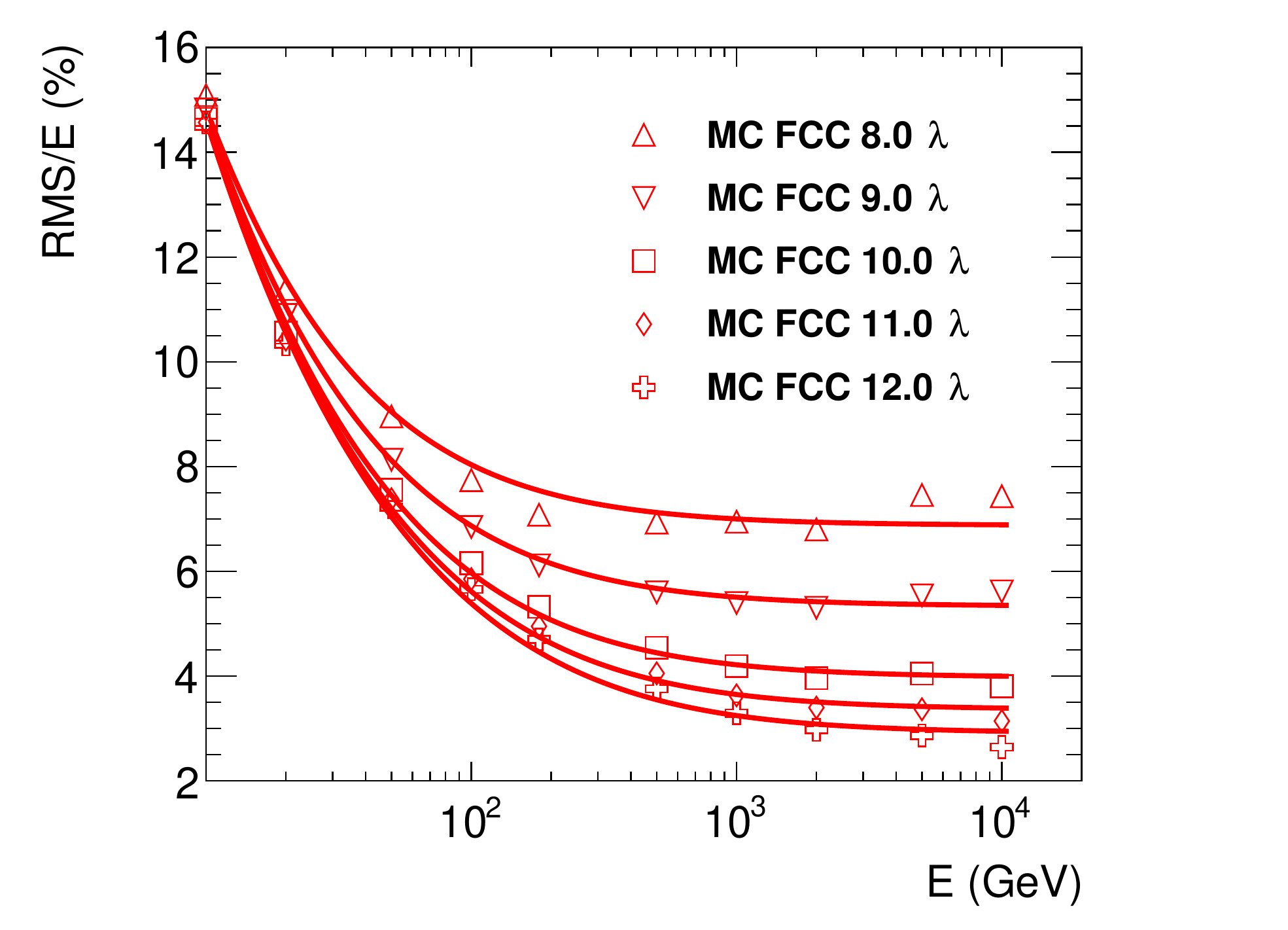}
 \caption{Single pion energy resolution extracted from a Gaussian fit (left) and as the RMS of the energy distribution (right). 
 Data corresponds to measurements by the ATLAS Tile calorimeter production modules at projective incident angle of $\eta$ = 0.35 
 corresponding to an effective depth of 7.9~$\lambda$ \cite{TileTestbeam}
 and from the studies on the prototype modules at the same angle with an effective depth of 9.5~$\lambda$ \cite{TilePrototypes}.
 \label{fig:Resolution}}
\end{figure}

Single hadron energy resolution is derived from the simulation of $\pi^{+}$ showers mentioned earlier. The energy resolution in a calorimeter is typically modeled by the following equation

\begin{equation}
\label{eq:reso}
\frac{\sigma(E)}{E} = \frac{a}{\sqrt{E}} \oplus \frac{b}{E} \oplus c
\end{equation}

where $a$ is the stochastic or sampling term, $b$ is the electronic noise term and $c$ is the constant term. 
All terms are added in quadrature. 
At high energies the constant term dominates but at intermediate energies the stochastic term might be not negligible. 
The electronic noise in the ATLAS Tile calorimeter is very small ($\sim$ 25~MeV/cell) 
and consequently is not considered in this simulation nor used in the fits to compare with test-beam data.
The resolution at different calorimeter depths shown in Figure~\ref{fig:Resolution} is obtained as a function of the energy 
of the incident pion extracted from the sigma of a Gaussian fit in $\pm2\sigma$ to the energy distribution (left) 
or as the RMS of that same distribution (right). 
Data is obtained from the studies on the production modules of the ATLAS Tile calorimeter 
at projective incident angle of $\eta$=0.35 for 7.9~$\lambda$ effective depth \cite{TileTestbeam} 
and from the studies on the prototype modules at the same angle with an effective depth of 9.5~$\lambda$ \cite{TilePrototypes}.
The results of the fits are summarized in Table~\ref{table:Resolution}.
The slightly more optimistic resolution obtained by the MC in comparison with the data 
(see Figure~\ref{fig:Resolution} and Table~\ref{table:Resolution})
can be explained by the absence of photo-electron yield, optics non-uniformities, calibration imperfections, and electronics noise in the new MC.
In addition in the MC the showers are shorter than in data (see Section~\ref{sec:Containment}).

\begin{table}[!ht]
\centering
\begin{tabular}{|c|c|c|c|c|c|c|}
\hline
& \multicolumn{4}{c|}{Simulation} & \multicolumn{2}{c|}{Data} \\
\hline
& \multicolumn{2}{c|}{Sigma} & \multicolumn{2}{c|}{RMS} 
& \multicolumn{2}{c|}{Sigma} \\
\hline
Depth ($\lambda$) 
& a (\%GeV$^{1/2}$) & c (\%) & a (\%GeV$^{1/2}$) & c (\%) 
& a (\%GeV$^{1/2}$) & c (\%) \\
\hline
8   & 41 & 5.0 & 42 & 6.9 & 52.9 & 5.7 \\
9   & 43 & 4.0 & 43 & 5.3 &   -  &  -  \\
9.5 & -  &  -  & -  &  -  &  54.5  &  4.0  \\
10  & 43 & 3.1 & 45 & 4.0 &   -  &  - \\
11  & 43 & 2.7 & 45 & 3.4 &   -  &  -  \\
12  & 43 & 2.4 & 45 & 2.9 &   -  &  -  \\
\hline
\end{tabular}
\caption{Fit values for data and simulation of energy resolution at different depths for both methods described in the text.
Data is extracted from the studies of production modules at an incident angle of $\eta$=0.35 for 7.9~$\lambda$ effective depth \cite{TileTestbeam} and from the studies on the prototype modules at the same angle with an effective depth of 9.5~$\lambda$ \cite{TilePrototypes}.
\label{table:Resolution}}
\end{table}


The energy resolution of pions with energy above 1 TeV degrades by a factor $\sim$2 
when passing from a calorimeter of 8~$\lambda$ to a long one of 12~$\lambda$. 
This degradation is mostly affecting the constant term, while the scaling term is rather insensitive to the depth of the calorimeter as is illustrated in Table~\ref{table:Resolution}.
The results show that the energy resolution achievable is $\sigma(E)/E = 45 \% / \sqrt{E} \oplus 3 \% $ in MC with a 12 $\lambda$ calorimeter.

\section{Conclusions}
\label{sec:Conclusions}

In this document, simulation results using a conceptual Geant4 detector based on the ATLAS Tile calorimeter have been presented.
The results obtained are used to explore the requirements for the Hadronic calorimetry at the FCC\_hh. 
Existing test-beam data have been used to validate the simulations at lower energies.
Simulations show how low energy tails below 3$\sigma$ are reduced by 70\% by going from a 8~$\lambda$ to a 12~$\lambda$ calorimeter.
According to the simulations of single particles used in this study the required calorimeter depth to contain single hadrons 
with energies above 1~TeV to 98~\% is $\sim$ 12~$\lambda$, this depth is also required to contain up to 98\% a $p_T$ = 40~TeV jet. 
A calorimeter with 12~$\lambda$ depth, inspired in the ATLAS-Tile concept will allow to keep the pion resolution to $\sim$ 3\% in the TeV region and get an overall energy pion resolution of 
$\sigma(E)/E = 45 \% / \sqrt{E} \oplus 3 \% $.

These studies were important to motivate a 12 $\lambda$ calorimeter for the central barrel calorimetry of FCC-hh baseline detector (ECAL+HCAL) , with the twin solenoid starting at 6~m radius. A stainless steel non-magnetic iron is needed in this case where the HCAL is inside the magnetic field. The advantage of using iron as an absorber versus more dense materials is to reduce the neutron content in the hadronic showers which consequently result in a faster response detector and reduction of the multiple scattering  of TeV muons before reaching the muon spectrometer.
The potential of the optics layout, with scintillating tiles parallel to the incoming particles at $\eta$ = 0 can be exploited in the future to obtain a much better perpendicular and transversal granularity than in ATLAS by using smaller photo-detectors as for example Si-PMTs.

\bibliographystyle{iopart-num}
\bibliography{Bibliography}

\end{document}